\documentclass[aps,prb,twocolumn,groupedaddress,showpacs,showkeys,floatfix]{revtex4}

\usepackage{amsmath}
\usepackage{amsfonts}
\usepackage{amssymb}
\usepackage{graphics}

\begin{document}

%\preprint{}

\title{Relaxation and reconstruction on (111) surfaces of Au, Pt, and Cu}

\author{\v Z.~Crljen}
\author{P.~Lazi\' c}
\author{D.~\v Sok\v cevi\' c}
\author{R.~Brako}
%\thanks{}
\affiliation{R. Bo\v skovi\' c Institute, P.O. Box 180, 10002 Zagreb,
     Croatia}

\date{\today}

\begin{abstract}
We have theoretically studied the stability and reconstruction of 
(111) surfaces of Au, Pt, and Cu. We have calculated the surface energy, 
surface stress, interatomic force constants, and other relevant quantities 
by \textit{ab initio} electronic structure calculations using the density 
functional theory (DFT), in a slab geometry with periodic boundary 
conditions. 
We have estimated the stability towards a quasi-one-dimensional 
reconstruction by using the calculated quantities as 
parameters in a one-dimensional Frenkel-Kontorova model. 
On all surfaces 
we have found an intrinsic tensile stress. This stress is 
large enough on Au and Pt 
surfaces to lead to a reconstruction in which a denser surface layer is 
formed, in agreement with experiment. The experimentally observed 
differences between the dense reconstruction pattern on Au(111) and a 
sparse structure of stripes on Pt(111) are attributed to 
the details of the interaction potential between the first layer of atoms 
and the substrate.
\end{abstract}


\maketitle

\section{\label{introduction}Introduction}

The properties of close-packed noble and transition metal surfaces
have been extensively studied in recent years.
The research has made great advances since the introduction
of new experimental techniques, in particular scanning tunneling
microscopy (STM), and the improvement of first-principle computational
methods based on the density functional approach. These surfaces
show a wide variety of behavior with respect to reconstruction,
preferred site and strength of chemisorption of reactive species,
adsorbate diffusion, etc.

Owing to the abrupt change of the electronic structure on metal
surfaces the first layer of atoms may have rather different
properties from the bulk. Often, a large intrinsic tensile surface
stress appears, which is the driving force leading to compressive
reconstruction on many close-packed metal surfaces. The stability
of  a particular surface is the result of the interplay of several
physical quantities characteristic of surface, such as
surface energy, surface stress, interatomic force constants,
etc.  The reconstruction is more likely to occur the larger
the intrinsic stress and the smaller the energy for sliding
the atoms of the first layers into positions out of registry
with the substrate.

In this paper we consider (111) surfaces of copper, platinum,
and gold.
The Au(111) surface reconstructs forming a uniaxially compressed
layer, which can be seen in STM experiments.\cite{Woll89,Harten85,
Repain00} The Pt(111)
reconstructs only at high temperatures,\cite{Sandy92,Grubel93}
or in the presence of saturated Pt vapour,\cite{Bott93,Hohage95}
and Cu(111) does not reconstruct.
We report on first-principle numerical calculations
of the properties of these surfaces using the density functional
theory (DFT). In Sec.~\ref{calculation} we first describe the generalities
of the computation procedure. We then calculate the important
physical quantities, such as the effective force constants 
in the first layer, the surface energies, and the intrinsic surface
stress. In Sec.~\ref{stability} we use these quantities
 in order to estimate 
the stability of the first layer of atoms with respect to a
quasi-one-dimensional reconstruction, which can be treated theoretically
by the Frenkel-Kontorova model. In the Sec.~\ref{conclusion} we discuss
the results.

\section{\label{calculation}Calculations of surface properties}

 The reconstructed phase of some close-packed metal surfaces may have 
a periodicity
which involves many surface atoms. For example, the reconstruction
of Au(111), one of the surfaces considered in this paper, involves
a uniaxially compressed first atomic layer with a period of
around  70~{\AA}. Full first-principle calculations with such large
supercells are not feasible. One instead evaluates the stability
by calculating the relevant quantities of the unreconstructed
surface, and uses them in phenomenological models of reconstruction.

There have been a number of papers in which the stability of 
(111) surfaces of noble metals was evaluated using the 
Frenkel-Kontorova (FK) model\cite{Frenkel38,Frank49,Zangwill96}
 of uniaxially compressive reconstruction. Ravelo and
 El-Batanouny (Ref.~\onlinecite{Ravelo89}, see also references therein) have 
constructed effective potentials between surface atoms and performed 
molecular-dynamics simulations of the reconstructed phase.
Mansfiled and Needs\cite{Mansfield90} have 
derived conditions of stability of (100) and (111) surfaces towards a 
compressive reconstruction in the FK model, and evaluated it for 
several metals. Apparently, their values of the parameters are not 
good enough, as they have found no reconstruction of 
Pt(111) and Au(111). In particular, the strength of the interatomic potentials 
seems to be overestimated.
Takeuchi et al.\cite{Takeuchi91}
 have constructed a two-dimensional potential between 
first- and second-layer atoms of a Au(111) surface using the results  
of first-principle calculations. They have calculated the reconstruction 
pattern using simulation techniques in a two-dimensional FK model. 
Using the two-dimensional potential in a molecular-dynamics simulation,
Narasimhan and Vanderbilt\cite{Narasimhan92}
 have concluded that the herringbone pattern
of the reconstruction stripes observed on Au(111) is favored by the 
long-range elastic interactions mediated by the substrate.
Recently, Narasimhan and Pushpa\cite{Narasimhan02}
 have studied a two-dimensional FK model of the
Pt(111) surface for which the parameters have been obtained 
from \textit{ab initio}
calculations. They have obtained simulated STM images of the reconstruction
stripes recombining and intersecting in various ways, similarly to the
structures observed in STM experiments.

The aim of this paper is to re-evaluate with highest accuracy
the properties 
of the surfaces which determine the stability towards reconstruction.
In this section we report on such calculations for the (111)
surfaces of Au, Pt, and Cu in the density functional theory (DFT)
approach.
In the following subsections we describe, in order, the generalities
of the numerical calculations, the calculation of surface energies,
surface stress, surface spring constants, and the calculation 
of energy required
to slide the complete first layer to various positions away
from the most stable one.
 In the next section we use these results as input
parameters in the (effectively one-dimensional) Frenkel-Kontorova
model of uniaxially compressive reconstruction.

\subsection{DFT calculations}

We have performed first-principle calculations of the electronic
structure of (111) surfaces within the density functional theory
approach, using the \textit{dacapo} program.\cite{Hammer99}
We have used the provided
ultrasoft pseudopotentials for the Perdew-Wang exchange-correlation
functional PW91 and the generalized gradient approximation (GGA).

We have first made calculations for bulk metal, in order to determine
the lattice constant at which the energy per atom is minimum,
using the primitive fcc unit cell. The lattice constant found
often differs slightly from the experimental value: for example,
our value is 4.00~{\AA} compared with the experimental value 3.92~{\AA}
for Pt, 4.18~{\AA} compared with the experimental value 4.08~{\AA}
for Au, and 3.66~{\AA} compared with 3.61~{\AA} for Cu. We have followed the
common practice using the computed value of the lattice constant
in susequent calculations, which ensures that no spurious bulk
stresses appear.

The surfaces were described by a slab of five or more hexagonal
layers of atoms. Since the calculation assumes periodicity in
all three dimensions, the metallic slabs were separated by typically
five layers of vacuum. In addition to ideal surfaces, we calculated
surfaces perturbed in various ways in order to deduce quantities
like surface stress and surface energy,  which can be used to
estimate the stability of the surface. More details are given
in the respective subsections.

In the case of a clean surface, the unit cell in the directions
along the surface plane $(x-y)$ consisted of one atom, and 56
$k$-points in the two-dimensional first Brillouin zone were used.
In some calculations described in the following subsections,
 e.g., when alternate rows of first-layer
atoms were displaced in opposite directions, in order to probe
the restoring forces within the first layer, unit cells with
four atoms in each plane were used, and the number of $k$-points
was reduced to 18. In the $z$-direction, perpendicular to the
layers, only $k=0$ was considered, consistent with the assumption
that the slabs, which repeat periodically because of the computational method,
did not couple to each other. In most calculations,
the two bottom layers were kept fixed at the bulk separation.
We used an energy cutoff for the plane-wave basis set of 340
eV, and the electronic occupation was smeared by a pseudothermal
distribution of $T = 0.2$ eV. We performed some checks with a
lower value, $T = 0.1$ eV, and found that the changes of the calculated
quantities were irrelevant.

Next, we calculated the structure of clean (111) surfaces. The
relaxation of the surface layers from truncated bulk positions
was found to be rather small for the three  metals considered.
For gold, the first and second layers relax less than 0.3\% 
of the interlayer distance in the $\left< 111\right> $
direction, with an energy gain of 1 meV, which is irrelevant
considering the accuracy of the calculation. The first platinum
layer relaxes outwards by about 1\%, with an energy
gain of 2.5 meV.
This unusual expansion of the interlayer distance has been observed
experimentally.\cite{Materer95}
On copper, the first three layers move inwards by less than 1\%,
and the energy gain is 4 meV.

\subsection{Surface energy}

We obtained the surface energy per atom as the difference of
the energy of the bulk and of a slab. In order to minimize the
influence of the computational details,  we calculated the energy
of the bulk using a six layer slab (i.e., twice the $abc$ stacking
of the $\langle 111\rangle$  fcc direction) with no
vacuum layers. After that, we changed the configuration by introducing four
layers of vacuum which created two (111) surfaces, and  calculated the
energy, allowing both surfaces to relax. The surface energies
$\gamma$, one half of the difference of the  energies
obtained in the two calculations, are given in the GGA column
in Table~\ref{parameters}. These values are consistently smaller
by about one third than those
calculated recently by Vitos et al.\cite{Vitos98} and Galanakis
et al.\cite{Galanakis02a,Galanakis02b} 
and various experimental values reported
therein. One possible source of the discrepancy are the different
pseudopotentials and exchange-correlation
functionals used. We therefore also report the surface energies
calculated using the LDA functional for the electronic densities
obtained in the GGA calculations, in the column labeled as LDA in 
Table~\ref{parameters}. These electronic densities are, of course, 
non-self-consistent with respect to the LDA functional. They
are always larger than the GGA results, in better agreement
with Refs.~\onlinecite{Vitos98,Galanakis02a,Galanakis02b}.
Another possible source of discrepancy is the small thickness
of the slab in our calculation. In the bulk calculation (i.e.,
6 layers without vacuum) we used only one $k$-point in the $z$-direction,
in order to keep the similarity with the surface calculations,
and consequently 
the long-distance contributions in the $z$-direction might not be
correctly taken into account. This can be investigated by using elementary
cells with more layers, i.e., repeating the $abc$ structure three
or more times, but we have not done such checks. Also, we used
the equilibrium lattice constants obtained self-consistently
in our GGA calculations which were  larger than the experimental
values and those obtained
by other authors from LDA calculations.

\begin{table}%[H] add [H] placement to break table across pages
\caption{\label{parameters}
Surface energy $\gamma$, surface stress $\tau$,
 and force constants between nearest
neighbors $k_{0}$, for (111) surfaces of Au, Pt and Cu. The two values of the 
surface energy $\gamma$ have been obtained from a fully self-consistent
DFT GGA calculation, and from applying the LDA functional to the same
electronic density, see text. The tree values of the force constant $k_{0}$
have been calculated from the bulk modulus $B$, $(k_{0B})$, and from
the forces obtained in DFT calculations in which the first-layer atoms were
slightly displaced "compressively" $(k_{0C})$ or "laterally" $(k_{0S})$,
see Fig.~\protect\ref{displacements}.
}
\begin{ruledtabular}
\begin{tabular}{ccccccc}
  & \multicolumn{2}{c}{$\gamma$ (eV/\AA$^2$)} & $\tau$ (eV/\AA$^2$)
       &  \multicolumn{3}{c}{$k_0$ (eV/\AA$^2$)}  \\
      \cline{2-3}\cline{5-7}
  & GGA  & LDA & &$ k_{0B}$ &$k_{0C}$ &$k_{0S}$   \\ \hline
Au  & 0.050  & 0.080 & 0.15 & 1.83 & 1.41 & 1.18  \\
Pt  & 0.084  & 0.123 & 0.34 & 3.12 & 2.23 & 2.55  \\
Cu  & 0.080  & 0.112 & 0.11 & 2.04 & 1.73 & 2.36 
\end{tabular}
\end{ruledtabular}
\end{table}
In the following section, when discussing the stability of the
surfaces, we shall use the LDA results, which seem to
agree well with the best values reported in the literature.

\subsection{Surface stress}

The surface stress can be found by considering the change of
energy when the lattice constant in the $x$ and $y$ direction (i.e.,
in the surface plane) is varied. In the bulk calculations the leading
correction to the energy when the lattice constant is varied
around the equillibrium value is quadratic by construction,
since we used the lattice constant corresponding to the energy minimum. 
Owing to the reduced coordination, the optimum interatomic distance
in the surface layer is smaller than the bulk lattice constant,
as already mentioned, causing an intrinsic tensile surface
stress.\cite{Ibach97,Haiss01}
Therefore, the surface contibution to the energy has
a leading linear term if the lattice constant in the $x$ and
$y$ direction is varied.

In order to find the intrinsic surface stress, we calculated the
energy of slabs consisting of six layers, as in the preceding
subsection, both unperturbed and with the unit cell compressed
or expanded along the coordinates $x$ and $y$ (i.e., in the directions
in the surface plane) for around 0.1\%, not allowing any additional
relaxation in the $z$ direction.
We also calculated similar configurations without vacuum layers,
in order to subtract the quadratic bulk term, as discussed earlier.
The diagonal element of the surface stress tensor $\tau$,
which is the relevant quantity for the surface reconstruction,
is
\begin{equation}
         \tau = \frac{\Delta E}{2\Delta A}, \label{tau}
\end{equation}
where $\Delta E$ 
is the change of energy when the
lattice constant is varied (after subtracting the quadratic
bulk term) and $\Delta A$ is the associated change of the surface
area. The numerical values are given in Table~\ref{results}. We have found that
in this case there is almost no change if the non-self-consistent
LDA values of the energy are used instead of the GGA values.

\subsection{Surface spring constants}

The simplest atomic-scale model of the lattice dynamics of fcc
metals is to assume central harmonic forces between nearest
neighbor atoms. In this model there is a universal scaling of
phonon spectra and different elements of the elasticity tensor,
which is obviously an oversimplification. Nevertheless, this
model is sufficient for the purpose of this work. The 
force constant $k_{0B}$ of the
nearest neighbor bond in the bulk can be found from the bulk
modulus\cite{Kittel71} $B = 1/3(C_{11}+2C_{12})$:
\begin{equation}
   k_{0B} = \frac{1}{2aB}, \label{spring}
\end{equation}
where $a$ is the lattice constant of the conventional fcc unit
cell.

However, the effective force constant between the atoms in the
first surface layer can differ from the value in the bulk,
since the lower coordination can substantially alter electronic
properties, and an ab-initio DFT calculation is necessary.
(Note that this force constant also contributes to the quadratic energy
term in the preceding subsection, where we did not consider the
possibility that it was modified within the first layer. However,
the possible error which this introduces in the calculation
of the surface stress is negligible.)

\begin{table*}%[H] add [H] placement to break table across pages
\caption{\label{results}
The parameters of the Frenkel-Kontorova model of uniaxial
compressive reconstruction. $A$ is the area per surface atom, $a$ is 
the distance between the fcc and bcc hollow site on the (111) surface. 
The other parameters were determined from the DFT
calculations as described in the text. These are the surface stress $\tau$, 
the non-self-consistent LDA value of the surface energy $\gamma_{LDA}$, the 
force constant $\mu = 3/2 k_{0C}$, and the average amplitude of the corrugated
potential of the second layer $W$. The quantities $\alpha$ and $\beta$ defined 
in Eq.~(\protect\ref{reconstruction}) determine the stability of the surface in the Frenkel-Kontorova
model.
}
\begin{ruledtabular}
\begin{tabular}{cccccccccc}
&  $A$ (\AA$^2$) & $a$ (\AA) & $\tau$ (eV/\AA$^2$) & $\gamma_{LDA}$
      (eV/\AA$^2$) & $\mu$ (eV/\AA$^2$) & $W$ (eV)
      & $\alpha$ & $\beta$ & Reconstruction \\ \hline
Au & 7.59& 3.41& 0.15& 0.080& 2.115& 0.038& 22.90& $-23.97$ & Compressive \\
Pt & 6.94& 3.27& 0.34& 0.123& 3.345& 0.061& 21.80& $-37.58$ & Compressive \\
Cu & 5.79& 2.99& 0.11& 0.112& 2.595& 0.056& 18.32& $-3.58$  & No
\end{tabular}
\end{ruledtabular}
\end{table*}

\begin{figure}
\resizebox{0.45\columnwidth}{!}{
\includegraphics{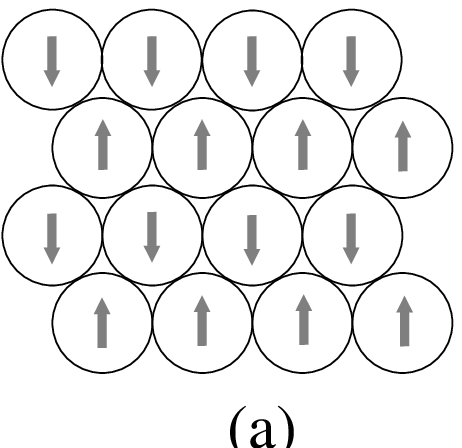} 
}
\resizebox{0.45\columnwidth}{!}{
\includegraphics{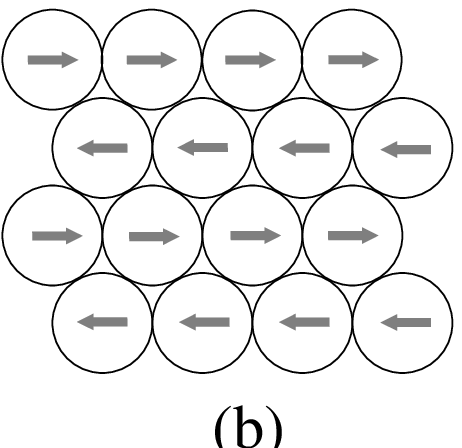}
}
\caption{\label{displacements}
The calculation of the force constants between the atoms in the surface layer
of (111) surface. The atoms in the top layer of the slab were displaced by
around 0.1\% of the interatomic distance, 
and the restoring forces were calculated.  Assuming 
that only central nearest-neighbor restoring forces exist, the force constants
are  $k_{0C}=(1/6) (F/{\delta y})$ for "compressive" displacements (a), and 
$k_{0S}= (1/2) (F/{\delta x})$ for "lateral" displacements (b).
}
\end{figure}
We made slab calculations (with relaxation turned off) in which
rows of atoms in the first layer were displaced by about 0.05 {\AA}
either in the direction of the rows or perpendicularly to
it, as shown in Fig.~\ref{displacements}. Unlike other calculations in this
paper, here we were not interested in the total energy, but
rather in the restoring forces appearing owing to the nonequlibrium
configurations. We analyzed the results assuming central harmonic
interactions (``springs'') between nearest neighbors, neglecting
the coupling to the second layer. Taking
into account the number of nearest neighbor bonds and the angles,
we calculated the force constants $k_{0S}$.
The results for both the bulk values obtained from Eq.~(\ref{spring}) and
the surface DFT calculations are shown in Table~\ref{parameters}.
 We note once
again that this approach does not attempt to give a complete
description of the lattice dynamical properties of the first
surface layer. In general, even if the interactions with subsurface
atoms are not taken into account, a set of force constants between
second and further neighbours and/or angular force constants
would be necessary in order to reproduce accurately the lattice dynamics,
i.e., the phonon
spectra. The present approach is a simplification to be used only in
an estimation of the stability of the surface layer. One
may further object that the reconstruction has a large
wavelength, while the displacements shown in Fig.~\ref{displacements}
correspond mainly to short-wavelength modes around the edge of the Brillouin
zone, and the estimated $k_S$ need not be the same. In our opinion,
the differences are small.

\subsection{Potential between the top layer and the bulk}

When reconstruction occurs, some first-layer atoms are displaced
to  energetically less favorable positions with respect to the
second layer. The last quantity which we  need  for an estimate
of the stability is the amount of energy
lost by the atoms when displaced to various nonoptimal positions.
Since the periodicity of the reconstruction pattern is large
compared with the substrate periodicity (e.g., by a factor of around
22 on Au(111)), the position of each subsequent atom along the
reconstruction direction with respect to the second layer changes
only slightly. A good estimate of the energy involved can be
obtained by considering structures in which all atoms in the
first layer are simultaneously displaced by the same vector.
 We performed DFT calculations of
such configurations, keeping the displacement in the $x-y$ plane
fixed and allowing the atoms to relax in the $z$ direction.

\begin{figure}
\resizebox{0.85\columnwidth}{!}{
\includegraphics{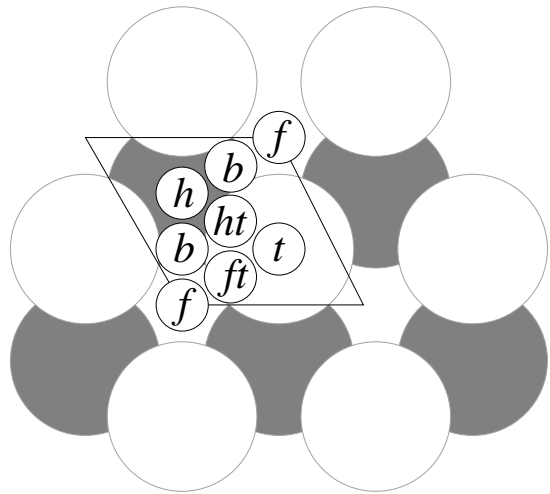}
}
\caption{\label{positions}
Various positions of the first-layer atoms (small circles) on a
(111) surface, relative to the second layer (large circles)
and the third layer (grey). In the unreconstructed phase, all
first-layer atoms are in the fcc positions ($f$). Upon reconstruction,
the atoms are found in various positions along the path
$f-b-h-b-f$.
}
\end{figure}
We first calculated the energy of the regular fcc configuration, denoted
by $f$ in figures and tables, which is energetically the most
favorable. Next, we considered the configuration with the first
layer atoms in the hcp hollows of the second layer ($h$), which
is also a local energy minimum. We furthermore calculated the
energy of some other configurations which are not local energy
minima, keeping the $x$ and $y$ coordinates fixed, so that the algorithm
for the atomic relaxation ignored the lateral forces acting
on the first-layer atoms. These are
(see Fig.~\ref{positions})
the ``bridge'' position ($b$) in which the first-layer atoms are
halfway between fcc and hcp hollows, the on-top configuration
($t$), the configurations ($ft$) and ($ht$)  halfway between ($f$)
and ($t$), and ($h$) and ($t$), respectively, etc. (Not all calculations
were performed for all three metals.) For the quasi-one-dimensional
compressive reconstruction, the relevant configurations are along
the path $f-fb-b-bh-h$ and back to $b$ and $f$. (The on-top configuration,
which is not occupied in the reconstruction, was calculated
with the aim to obtain a better insight into the difference
between various metals.)
In practice, the calculations were performed so that initially, all
layers except the top one were kept fixed at the bulk configuration,
the top layer had fixed $x$ and $y$ coordinates, and the $z$ coordinate
was allowed to relax. In a second step, the bottom two
layers were kept fixed, the intermediate layers were allowed
to relax in all directions, and the top layer was allowed to
relax in the $z$ direction only. The second step produced only minor
changes for the symmetrical configurations ($f$), ($h$), ($t$), and
($b$), but the nonsymmetric configurations changed
significantly, as the atoms in the second layer (and to a lesser
degree in deeper layers) relaxed laterally under the force exerted
by the first-layer atoms.

\begin{figure}
\resizebox{\columnwidth}{!}{
\includegraphics{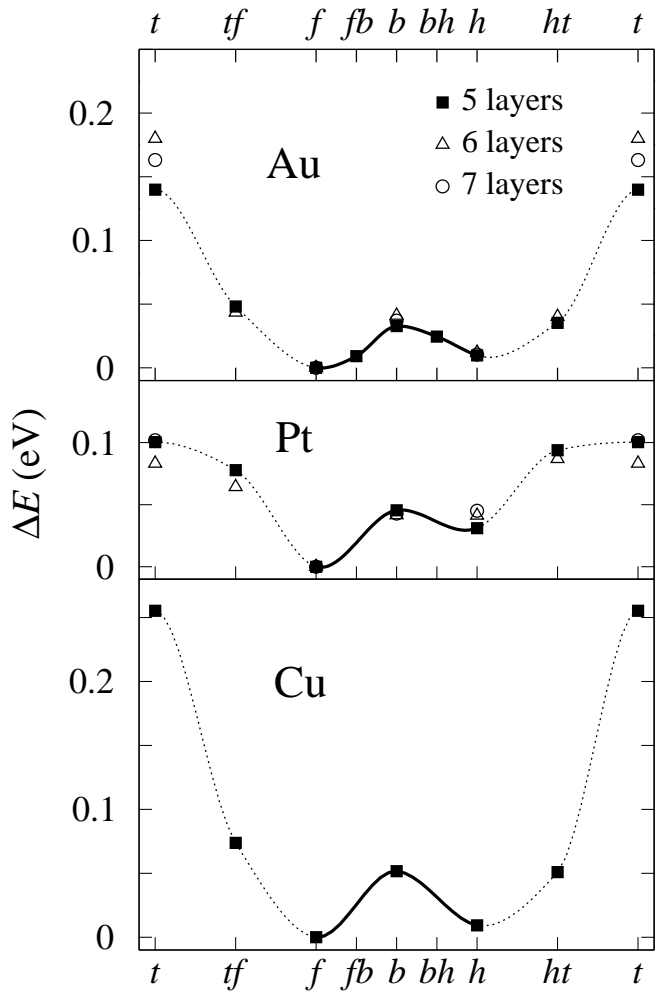}%
}
\caption{\label{Au}
The energy needed in order to move the complete first atomic
layer away from the optimum fcc into other configurations.
The various positions are depicted in Fig.~\protect\ref{positions},
and the energies are expressed in eV/atom.
}
\end{figure}
In Fig.~\ref{Au} we show the calculated energies, measured with respect
to the energy of the fcc configuration, which was the
lowest for all three metals. The ``bridge'' position, 
midway between fcc and hcp hollows,
is approximately a saddle point. All points were calculated
for five layers (first one fixed in $x$ and $y$, two free, two fixed),
and some points were also calculated for six and seven layers. It
was found that the energies  changed by several meV, but the
qualitative behavior remained similar, and the estimates of
stability towards reconstruction
in the following section are not affected. 
 Generally, the energies of symmetric configurations
($h, b, t$) increase with increasing number of layers, while
those of nonsymmetric configurations, like ($fb$) or ($ht$), decrease.
This can be attributed to the fact that in the latter case
there are more atomic
layers which are free to relax laterally, leading to a more
complete relaxation.

The first-layer atoms in positions other than the regular fcc one 
($f$) are also protruding further out.
For all three metals, the values are similar, namely, around
0.1~{\AA} for the bridge ($b$), 0.02~{\AA} for the hcp ($h$),
and $0.3-0.4$~{\AA} for the on-top position ($t$).

The relevant quantity for the stability estimate is the average
 energy $W$ along the path $f-b-h-b-f$, in the approximation of
making a Fourier expansion and
 keeping only the two lowest terms: 
\begin{equation}
W = \Delta E(b) + \Delta E(h) / 2. \label{W}
\end{equation}

\section{\label{stability}Stability and reconstruction}

 In this section we use the calculated quantities to discuss
 the stability of the surfaces of platinum,
 gold and copper. 
 We first introduce the one-dimensional
 Frenkel-Kontorova model, and then discuss, in order, each
 of the considered surfaces, calculating the stability criteria,
 and comparing the results with the known experimental findings.

\subsection{Reconstruction of (111) surfaces}

As already said, the electronic structure of closed-packed metal
 surfaces is very different from that  in the bulk, with an abrupt
 decrease in density of conduction electrons. The consequences
 may be a change of the length of the bonds between surface atoms,
 occasionally accompanied by a reconstruction which involves
 a change of the number of atoms in the first surface layer.

The reconstruction of the fcc (111) surface involves a large number
 of atoms in inequivalent positions with respect to the underlying
 layer of atoms.  In scanning tunneling microscopy experiments,
 the reconstruction of the Au(111) surface
 can be clearly seen as bright stripes,\cite{Woll89,Repain00,Bach97}
 owing to the increased
 height of atoms which are out of registry with the second layer.
  On a larger scale, the stripes form a herringbone pattern,
 as the quasi-one-dimensional reconstruction proceeds along three
 equivalent directions on the surface. A typical size of the
 reconstructed Au(111) supercell is $70{\textrm \AA} \times 280{\textrm \AA}$.
 The reconstruction of the Pt(111) surface is similar, but occurs only
 at high temperatures\cite{Sandy92,Grubel93}
or in the presence of saturated Pt vapour.\cite{Bott93,Hohage95}
STM micrographs
 show that the stripes do not form continuous patterns, but are
 instead well separated, with various types of intersections.\cite{Hohage95}
 The stripes in both systems consist of quasi-one-dimensional
 compressive reconstruction\cite{Woll89,Harten85}
 and can be treated as solitonic solutions
 of the Frenkel-Kontorova model.\cite{Takeuchi91,Narasimhan92} The large-scale
 two-dimensional structure of the reconstructed phase  depends
 upon details of the  interactions, and will not be discussed
 here.

\subsection{The Frenkel-Kontorova model}

The one-dimensional  Frenkel-Kontorova model\cite{Frenkel38,Frank49,Zangwill96}
 consists of a linear chain of atoms, subject to two competing
 interactions with different intrinsic periodicities. The atoms
 interact via a nearest-neighbor harmonic coupling with a preferred
 lattice constant $b$. In addition, there is an external periodic
 potential (i.e., the potential of the second atomic layer)  of
 a periodicity $a$. Depending on the strength of the external potential
 and the magnitude of the spring constant $\mu$, various stable
 solutions are possible. At small ``pressures'' (i.e., small difference
 of $a$ and $b$, small $\mu$) the external periodic potential dominates
 and all atoms are in potential minima, forming a commensurate
 phase. As the ``pressure'' increases, the natural periodicity
 of the atomic chain becomes more important. If, say, 
 $b<a$, solitons appear in which the atoms are closer to each other,
 thus gaining energy from the harmonic interaction but paying
 the cost of increased potential energy in the external potential.
 Finally, when the external potential is weak compared with the
 elastic energy needed to stretch the atomic chain, the atoms
 follow the periodicity $b$, forming an incommensurate phase.

Mansfield and Needs\cite{Mansfield90}
have applied the Frenkel-Kontorova
 model to the reconstruction of (111) surfaces of the fcc metals.
 They have found that the relevant quantity is
\begin{equation}
P=\frac{ A(\gamma-4\tau/3)}{ \frac{2}{\pi}\sqrt{2 \mu W a^2}}
= \frac{W\beta}{W\alpha}, \label{reconstruction}
\end{equation}
where the parameters of the Frenkel-Kontorova model have already
 been expressed in terms of the physical properties of real (111)
 surfaces. Thus, $\gamma$ is the surface energy,
 $\tau$ is the surface stress, $\mu$
 is 3/2 of the surface force constant, and $W$ is the average potential
 energy. (The factors 4/3 in the stress term and 3/2 in the definition of
$\mu$ appear because the path $f-h-f$ is not straight.) 
$a$ is the distance from the fcc hollow site to the nearest
 hcp hollow site on the (111) surface, and $A$ is the surface area
 per atom. These quantities are the same (or closely related
 to) those calculated in the preceding section for real metal
 surfaces. There is no reconstruction for $|P|<1$,
 the reconstruction is compressive for $P<-1$ 
 and expansive for $P>1$. The quasicontinuum
 approach is valid if the magnitudes of the dimensionless quantities
 $\alpha$ and $\beta$ defined in Eq.~(\ref{reconstruction})
 are large, say larger than 5.

In the following subsections we apply this analysis to the (111)
 surfaces of gold, platinum, and copper. The relevant quantities
 derived in Section~\ref{calculation} are summarized in Table~\ref{results},
 and the resulting $\alpha$ and $\beta$ are given.

\subsection{Au(111)}

In our calculations the first layer of an ideal Au(111) surface relaxes
 outwards by about 0.005~{\AA}, less than 0.3\%, which is at the
 limit of accuracy of the calculations. The energy per surface
 atom of the hcp configuration is higher than that of the ideal
 (fcc) structure by around 10 meV per atom, and that of the
 bridge by 33 meV (Fig.~\ref{Au}). This is in excellent agreement with
 the calculations of Takeuchi et al.\cite{Takeuchi91} and
 Galanakis et al.,\cite{Galanakis02a,Galanakis02b} who used a density
 functional approach with a mixed basis set. These energies are
 the smallest of all metals considered here. The surface energy
 is also small, and owing to a moderately large tensile surface
 stress we obtain that $\beta$ is slightly larger
 than $\alpha$, indicating that the surface reconstructs.
 Experimentally, the surface appears particularly prone to reconstruction,
 and a dense pattern of stripes of $23\times\sqrt3$ reconstruction
 is observed at room temperatures with STM.\cite{Woll89,Bach97,Repain00}

\subsection{Pt(111)}

 Our calculations show that the first atomic layer on Pt(111)
 relaxes outwards by 0.023~{\AA} or 1\%. This somewhat exceptional
 behavior has also been found in other calculations and confirmed
 experimentally.\cite{Materer95}  It has been attributed to an
 outward pointing electrostatic force on the positively charged
 atoms of the first layer owing to the kind of spill-out of electrons
 away from the surface.\cite{Needs87} The energy of the bridge configuration
 is around 46 meV larger than that of the regular fcc surface.
 The energy of the hcp configuration is only slightly smaller
 than the bridge (even less so in calculations with 6 and 7 layers),
 which is different from the other surfaces considered. The force
 constant $k_{0C}$ calculated assuming a ``longitudinal'' displacement
 of rows of first-layer atoms is reduced by about 30\%  compared
 with that derived from the bulk compressibility. The intrinsic
 tensile surface stress is large, more than twice larger
 than in the other two metals considered. Although the surface
 energy is also quite large, the quantity $\beta$
 is much larger than $\alpha$, predicting a strong
 tendency to reconstruct.

 Experimentally, the reconstruction of Pt(111) is observed only
 at high temperatures or in the presence of saturated platinum
 vapor.\cite{Sandy92,Grubel93,Bott93,Hohage95}
 The nature of the reconstruction is similar to that
 of Au(111), but the stripes of compressive ``solitons'' do not
 form a dense pattern. Instead, they are  sparse, and occasionally
 intersect in several distinct ways.\cite{Hohage95,Narasimhan02}

\subsection{Cu(111)}

Unlike the other two metals considered, we find that the first
 layer on an ideal (fcc) Cu(111) surface relaxes inwards by 0.014~{\AA}.
The contraction of the interlayer distance is usual for close
 packed surfaces of many transition metals. The hcp configuration
 is a clear local energy minimum, and the energy of the on-top
 configuration is found to be exceptionally large at about 250
 meV. The intrinsic surface stress is again tensile, but rather
 small. The calculated $\beta$ is clearly smaller
 than $\alpha$, indicating that the surface does
 not reconstruct. Indeed, no reconstructruction has been observed
 experimentally.

\subsection{Discussion}

The use of the stability condition of (111) surfaces derived in the
one-dimensional Frenkel-Kontorova model, Eq.~(\ref{reconstruction}),
gives correct 
qualitative answers, namely, that the surfaces of gold and platinum 
reconstruct compressively and the surface of copper does not. 
From a quantitative point of view, the agreement between theory 
and experiment is less satisfactory.
We find that the condition for reconstruction, 
$|\beta| > \alpha$, is barely satisfied for the Au(111) surfaces,
while in experiments a dense pattern of reconstruction stripes is
always found on this surface. On the other hand, we obtain that
the condition of Eq.~(\ref{reconstruction}) is amply fulfilled for the Pt(111) 
surface, while
experimentally, reconstruction is observed only under favorable 
thermodynamic conditions, at high temperatures, or in the presence
of Pt vapor.
It has already been noticed that calculations predict only
a marginal tendency for reconstruction on Au(111), at variance
with experiment. It has been suggested that the ordering
of the reconstruction stripes into a 
herringbone pattern is indeed an essential contributing factor 
to the stability of the reconstructed phase.\cite{Bach97,Narasimhan92}
The theoretical overestimate of the tendency to reconstruct 
in the case of 
Pt(111) may be due to the large value of the surface energy,
which makes it unfavorable to bring additional atoms to the
surface once the surface stress has been reduced by the formation
of  a few sparse reconstruction stripes.
For a more complete treatment, a two-dimensional Frenkel-Kontorova
model with a more realistic potential, calculated in a region 
around the minima, should be used.

The approach used in this paper can also give relevant parameters
for other physical properties. We have calculated relaxations around 
adatoms of the same species on the surfaces considered in this
paper, and found that the force--relaxation ratio is consistent
with the model of elastic constants calculated in 
Section~\ref{calculation}. Surprisingly, in our calculations
(one adatom per four surface atoms) the adsorption into a hcp
hollow site was slightly more favorable than into a fcc hollow 
site for all three metals considered. 
We have not found any conclusive  experimental data about
this property.

The approach can also be applied to relaxation around defects 
and chemisorbates of different species, energetics of self- and 
heterodiffusion, and other properties of (111) surfaces.

\section{\label{conclusion}Conclusions}

Using the density functional theory approach we have calculated
 key properties of (111) surfaces of several metals, 
and evaluated the stability towards reconstruction
 into a uniaxially compressed reconstructed layer, using
 a one-dimensional quasicontinuum Frenkel-Kontorova model.
The intrinsic surface stress is tensile for all surfaces, i.e., the
 atoms in the first layer prefer a denser packing than dictated
 by the potential of the second layer, which reflects the bulk
 periodicity. On the opposite side of the energy balance is the
 energy required to bring an extra atom into the surface layer,
 and the energy lost owing to the fact that surface atoms in a reconstructed
 phase are no longer in the minima of the potential of the second
 layer. 
 We have found that Pt(111) reconstructs owing to a large intrinsic
 tensile surface stress. The stress on Au(111) is considerably
 smaller, but the surface energy is also smaller and the reconstruction
 criterion is satisfied. The intrinsic surface stress in Cu(111)
 is somewhat smaller than that in Au(111), but other quantities
 are unfavorable, and the surface does not reconstruct, in agreement
 with experimental observations. Experiments show that Au(111)
 has the largest tendency to reconstruct, and dense reconstruction
 patterns are observed using STM, while Pt(111) reconstructs only
 under favorable conditions, and only a small fraction of the
 surface is reconstructed. This is the opposite from what our
 reconstruction criteria suggest, where Pt(111) is much deeper
 in the reconstruction regime. The reason may be further stabilization
 of the reconstruction of the gold surface by the formation of
 a two-dimensional herringbone structure,\cite{Bach97,Narasimhan92,Bulou02}
  which is not
 included in the one-dimensional model used in the derivation.

\begin{acknowledgments}
This work was supported by the Ministry of Science and Technology
of the Republic of Croatia under contract No. 0098001.
One of us (R.~B.) acknowledges helpful discussion with 
Dr. S.~Narasimhan.
\end{acknowledgments}


\end{document}